\documentclass[twoside]{ilcws08} 
\usepackage[latin1]{inputenc} 
\usepackage[dvips]{graphicx,epsfig,color} 
\usepackage{wrapfig,rotating} 
\usepackage{amssymb,amsmath,array}

\def\to{\rightarrow}
\def\ie{{\it i.e.}}
\def\eg{{\it e.g.}}

\begin{document} 
\title{
SUSY Without Prejudice at Linear Colliders}  

\author{Thomas G. Rizzo  
\thanks{Work supported in part by the Department of Energy, Contract 
DE-AC02-76SF00515} 
\vspace{.3cm}\\ 
SLAC National Accelerator Laboratory \\ 
2575 Sand Hill Rd., Menlo Park, CA, USA }

\maketitle

\begin{abstract}
We explore the physics of the general CP-conserving MSSM with Minimal Flavor Violation, 
the pMSSM. The 19 soft SUSY breaking parameters are chosen so to satisfy all  
existing experimental and theoretical constraints assuming that the WIMP is the 
lightest neutralino. We scan this parameter space twice using both flat and log 
priors and compare the results which yield similar conclusions. Constraints from 
both LEP and the Tevatron play an important role in obtaining our final 
model samples. Implications for future TeV-scale $e^+e^-$ linear colliders(LC) 
are discussed.
\end{abstract}

The MSSM with R-parity conservation is the prototypical SUSY model. Although it has many 
nice features it is difficult to study in detail as SUSY breaking leads to over 120 
{\it a priori} unknown soft-breaking parameters. The usual approach to this problem in 
phenomenological studies is to examine only one or a few of the usual SUSY breaking 
mechanisms, \eg, mSUGRA, GMSB or AMSB which have only a few associated parameters. One may 
ask whether a study of these models really tells us all we need to know about the MSSM; is 
there nothing more? How much do we really know about the MSSM?

In recent work, we have attempted to go beyond these studies{\cite {us}} by examining the 
general CP-conserving MSSM with Minimal Flavor Violation assuming that ($i$) the first 2 
generations of sfermions are 
degenerate with negligible Yukawa couplings, ($ii$) the neutralino is the LSP with the 
WMAP dark matter constraint used only as an upper bound. This leaves us with {\it only} 19 
independent, real, SUSY/weak-scale Lagrangian parameters to consider. These include the gaugino 
masses, $M_{1,2,3}$, the  Higgsino mixing parameter, $\mu$, the ratio of the Higgs vevs, $\tan \beta$, 
the mass of the pseudoscalar Higgs boson, $m_A$, and the 10 squared masses of the sfermions 
(5 for the assumed degenerate first two generations and a separate 5 for the third generation). 
Finally, due to the small Yukawa couplings for the first two generations, independent 
$A$-terms are only phenomenological relevant for the $b,t$ and $\tau$ sectors.
In the analysis presented here we will make two independent parameter scans: for
the first scan of $10^7$ parameter points we assume that all of the sfermion mass parameters lie in the 
range 0.1-1 TeV, $1 \leq \tan \beta \leq 50$, $|A_{b,t,\tau}| \leq 1$ TeV, $43.5 \leq m_A \leq 1000$ GeV, 
$0.05 \leq |M_{1,2},\mu|\leq 1$ TeV and $0.1 \leq M_3\leq 1$ TeV. Note the absolute value signs in these 
quoted ranges as we will allow the soft-breaking parameters to have arbitrary sign. 
For this first scan, which generates SUSY spartners with light to moderate 
masses, we will assume flat priors, \ie, we assume that the parameters values are chosen {\it uniformly} 
throughout their allowed ranges. For our second scan, for which will assume a set of log priors for the 
mass parameters, we will modify the allowed soft parameter ranges as follows: all the sfermion mass 
parameters will be assumed to lie in the range 0.1-3 TeV, $1 \leq \tan \beta \leq 60$, 
$0.01 \leq |A_{b,t,\tau}| \leq 3$ TeV, $43.5 \leq m_A \leq 3000$ GeV, $0.01 \leq |M_{1,2},\mu|\leq 3$ TeV, 
and $0.1 \leq M_3\leq 3$ TeV. This expanded range will allow us access to both very light as well as  
heavy sparticle states. Since the goal of this second scan is to make 
contrasts and comparisons to the first, flat prior study, we will only generate a set of $2\cdot 10^6$ 
parameter points to examine in this case.  

Once the model points are generated we pass them carefully through an extensive list of experimental  
constraints arising from precision measurements(\ie, $\Delta \rho$ and the invisible width of the $Z$), 
flavor physics(\ie, meson-antimeson mixing, $b\to s\gamma$, $B\to \tau \nu$ and $B_s\to \mu^+\mu^-$), the muon $g-2$, 
5-year WMAP data, direct detection searches(from CDMS, XENON10, DAMA and CREST), as well as Higgs boson and SUSY particle direct 
searches at both LEP and the Tevatron{\cite {us}}. Many of the caveats associated with these collider searches, 
which are usually ignored, have been included in the present analysis. An example of this is the search for right-handed 
sleptons at LEPII; usually, these limits are quoted as being $\sim 100$ GeV. However, if the mass splitting of the 
slepton with the LSP, $\Delta m$, is sufficiently small the resulting leptons would be too soft to be observable above SM 
backgrounds even in an $e^+e^-$ environment. Thus, lighter sleptons may actually exist with small values of $\Delta m$ 
that have escaped detection. This effect is indeed shown on the figures displaying the combined LEPII limits. A similar 
effect can also occur for charginos, squarks and gluinos and they, too, may be lighter than is commonly believed if their 
mass splitting with the LSP is sufficiently small. Such states do occur in our final model samples.   

It is important to note that the Tevatron plays a very important role at obtaining our final model set, in 
particular searches for stable charged particles, multijets plus missing energy and trilepton plus missing energy 
final states. These constraints, which we have analyzed at the level of fast detector simulations, have not been 
employed in previous SUSY analyses. We find that only $\sim 68.5(3.0)$k models survive this procedure in the case 
of the flat(log) priors.  

\begin{figure}[htbp]
\centerline{
\includegraphics[width=13.0cm,angle=0]{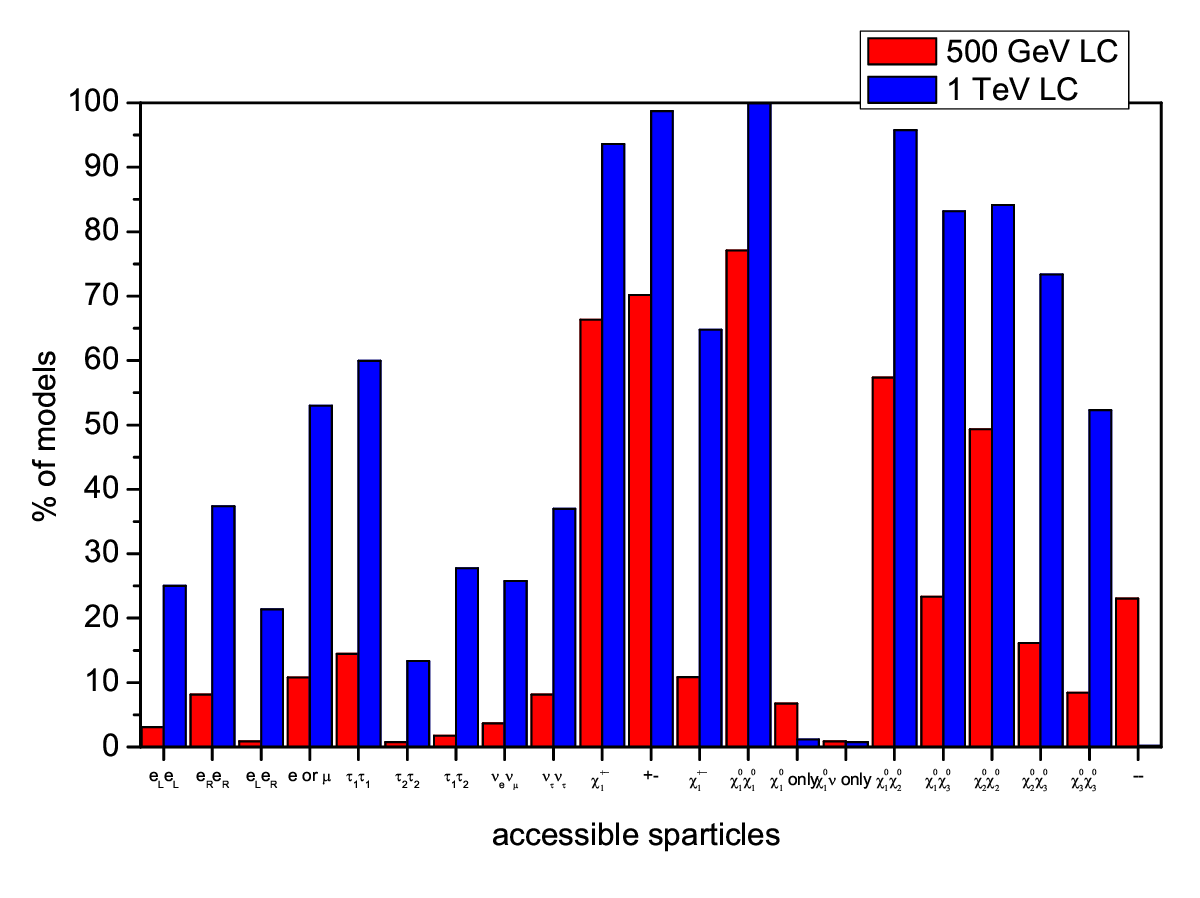}}
\vspace*{0.1cm}
\centerline{
\includegraphics[width=13.0cm,angle=0]{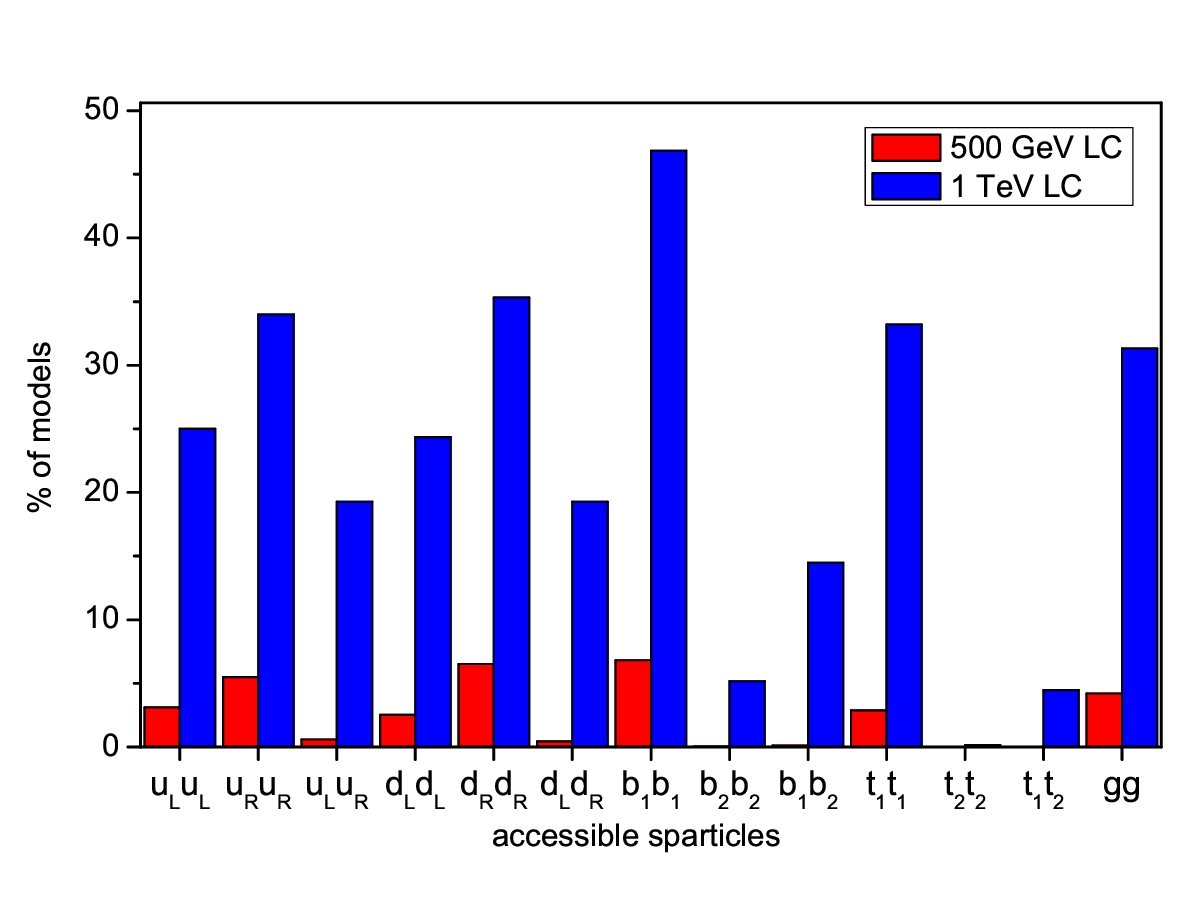}}
\vspace*{-0.3cm}
\caption{Fraction of the full model sample with kinematically accessibility of SUSY partners at a 500 GeV and a 
1 TeV LC assuming flat priors.}
\label{Fig1}
\end{figure}
\begin{figure}[htbp]
\centerline{
\includegraphics[width=13.0cm,angle=0]{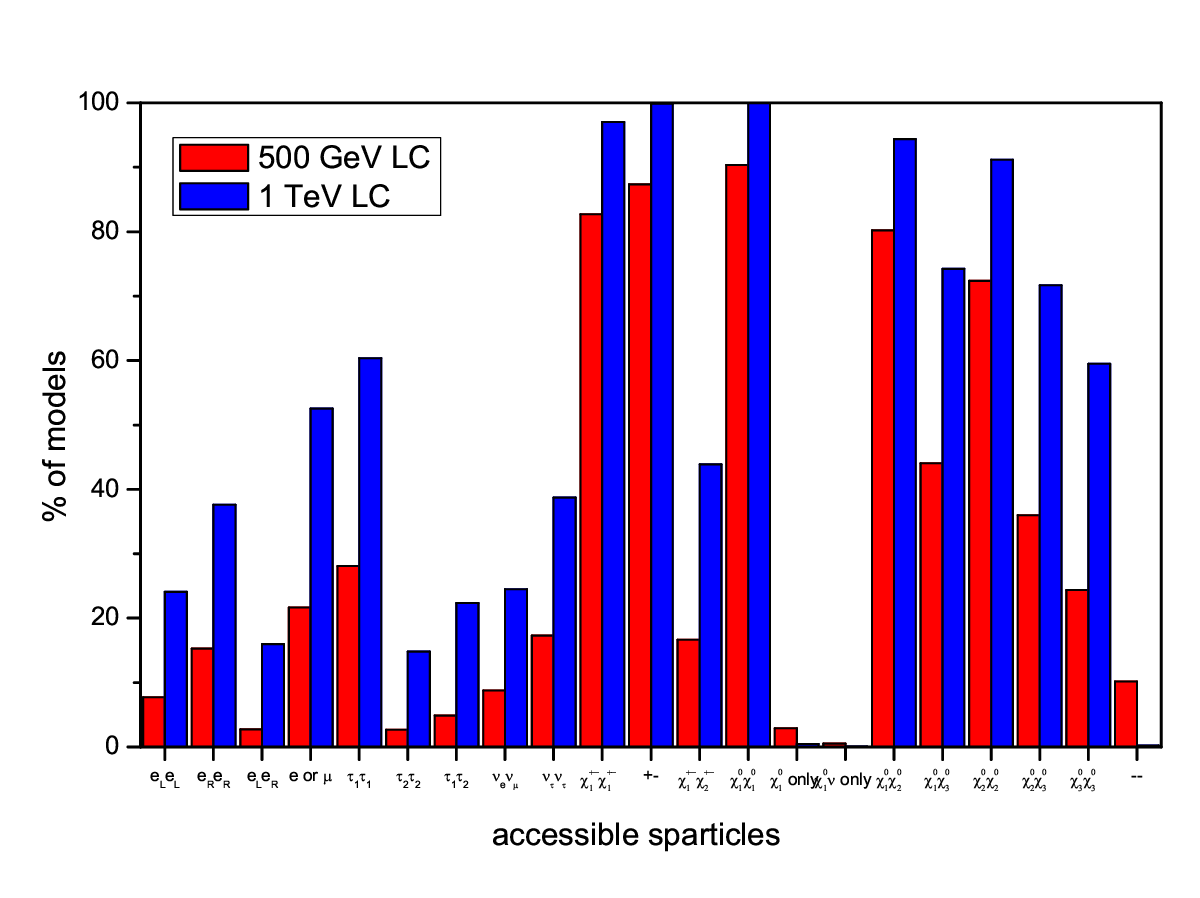}}
\vspace*{0.1cm}
\centerline{
\includegraphics[width=13.0cm,angle=0]{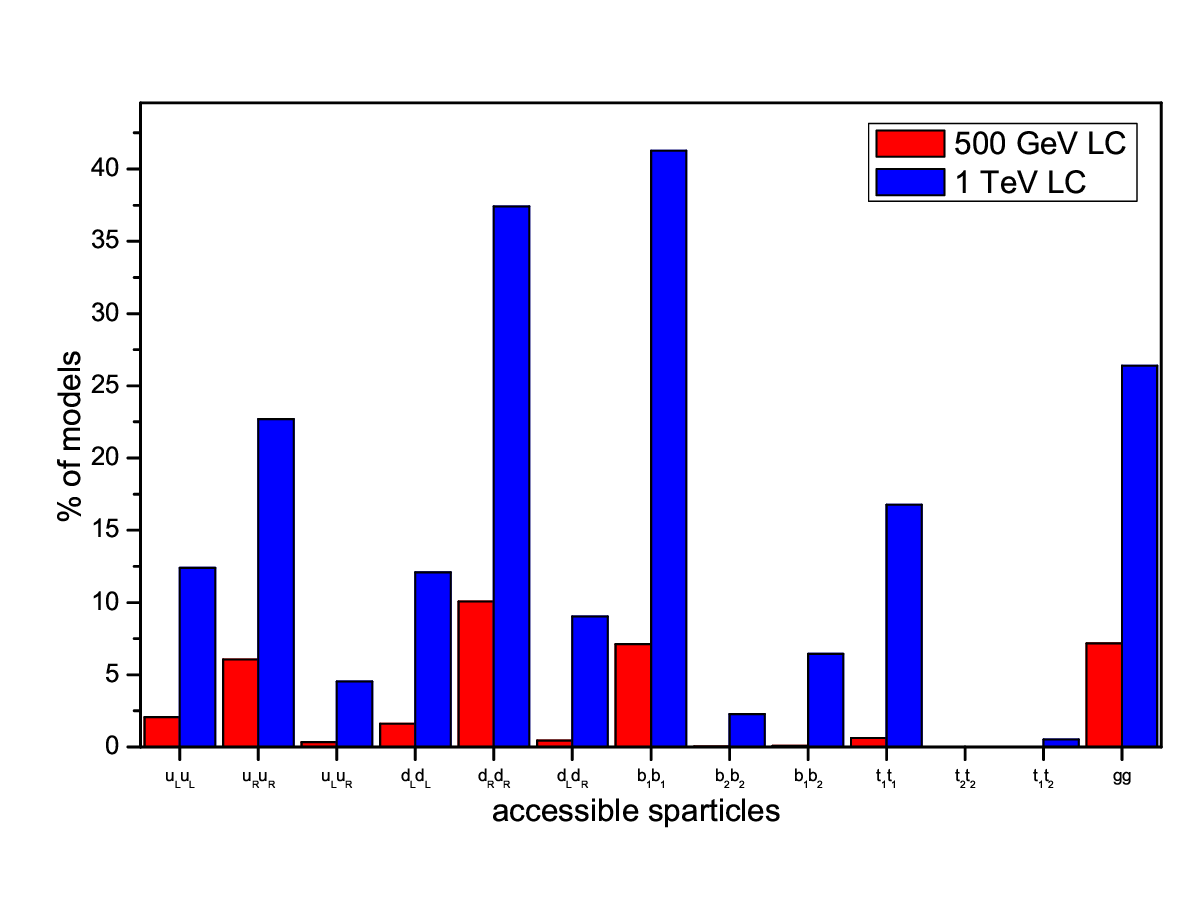}}
\caption{Same as the previous figure but now for log priors.}
\label{Fig2}
\end{figure}

For LC perhaps the most important result of this analysis is the sparticle spectra and whether or not a given sparticle 
is kinematically accessible at LC energies. Figs.~\ref{Fig1} and ~\ref{Fig2} show this accessibility at 
collision energies of 
either 500 GeV and 1 TeV for both sets of priors for many possible final state spartner pairs. Here, \eg, `$u_Lu_R$' 
labels the production of the $\tilde u_L \tilde u_R^*+h.c.$ associated production 
final state. Several things are to be noted from these figures: ($i$) The LSP is 
generally quite light as is $\tilde \chi_1^\pm$ since wino-like and Higgsino-like LSP spectra are the most common in 
both samples. ($ii$) The $\tilde \chi_2^0$ is also often very light as would happen in Higgsino-like scenarios. 
Note that  $\tilde \chi_1^\pm$ (and sometimes  $\tilde \chi_2^0$ as well) is often nearly degenerate with the LSP 
leading to long-lived charged particle collider signatures. 
($iii$) Charged sleptons may be accessible in some cases at 500 GeV with the lightest stau being the most likely 
possibility. ($iv$) There is a reasonable probability that light squarks may also be accessible at 500 GeV; this 
possibility has {\it not} been well-studied at LC since it has generally been believed that this possibility is 
excluded by Tevatron searches. Such searches, however, assume an mSUGRA-like mass spectra which produce large missing 
energy and multi-jet final states. In the cases we find, light squark production fails to pass the required Tevatron 
search cuts at significant rates to be detected, evading such searches. ($v$) There is a reasonable probability that 
a light Higgs exists with a mass below the LEPII SM Higgs bound of 114 GeV due to a reduced $ZZh$ coupling and/or the 
possibility that $h\to 2\tilde \chi_1^0$ is allowed with a significant rate. The former generally occurs in the 
`non-decoupling' region when $M_A$ is relatively small. These possibilities has not been well explored within the pure 
MSSM context. 

\begin{figure}[htbp]
\centerline{
\includegraphics[width=14.0cm,angle=0]{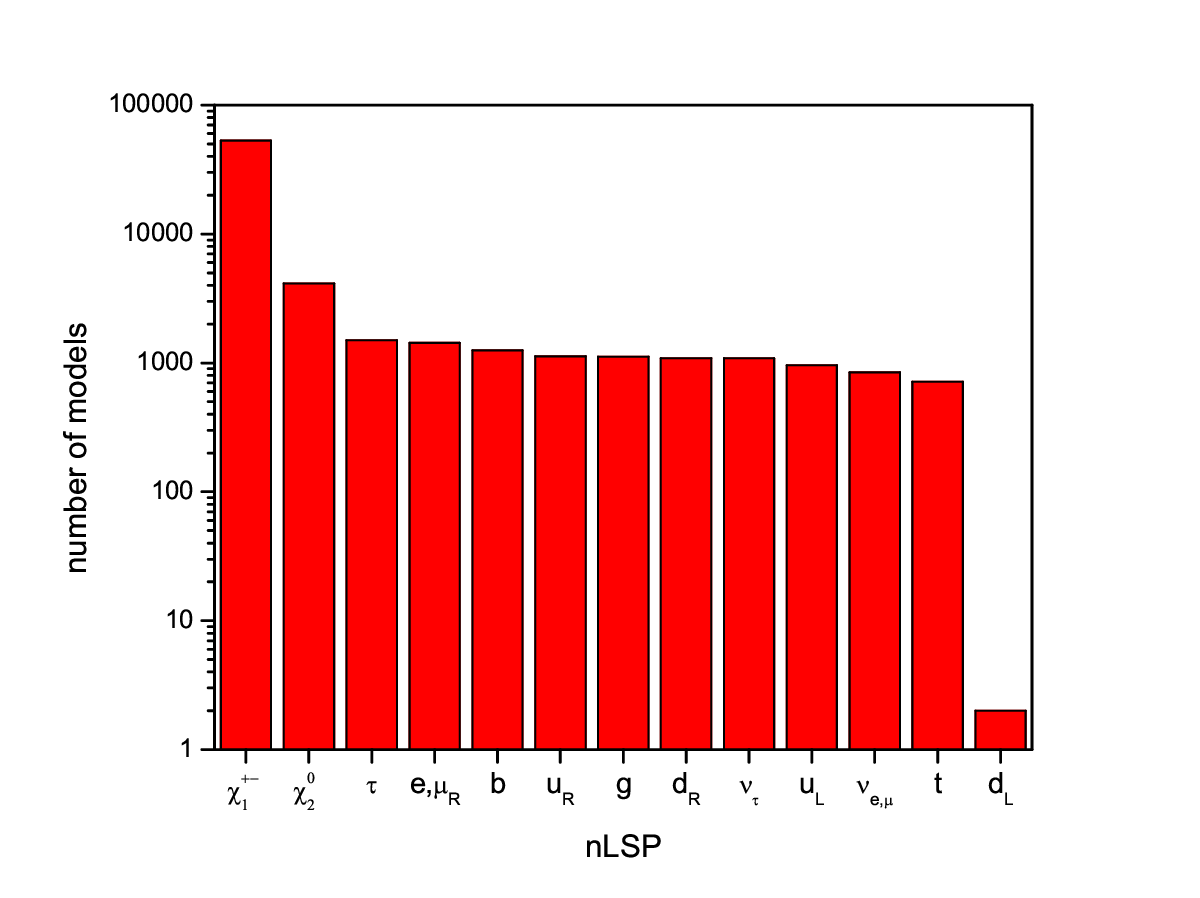}}
\vspace*{-0.1cm}
\centerline{
\includegraphics[width=14.0cm,angle=0]{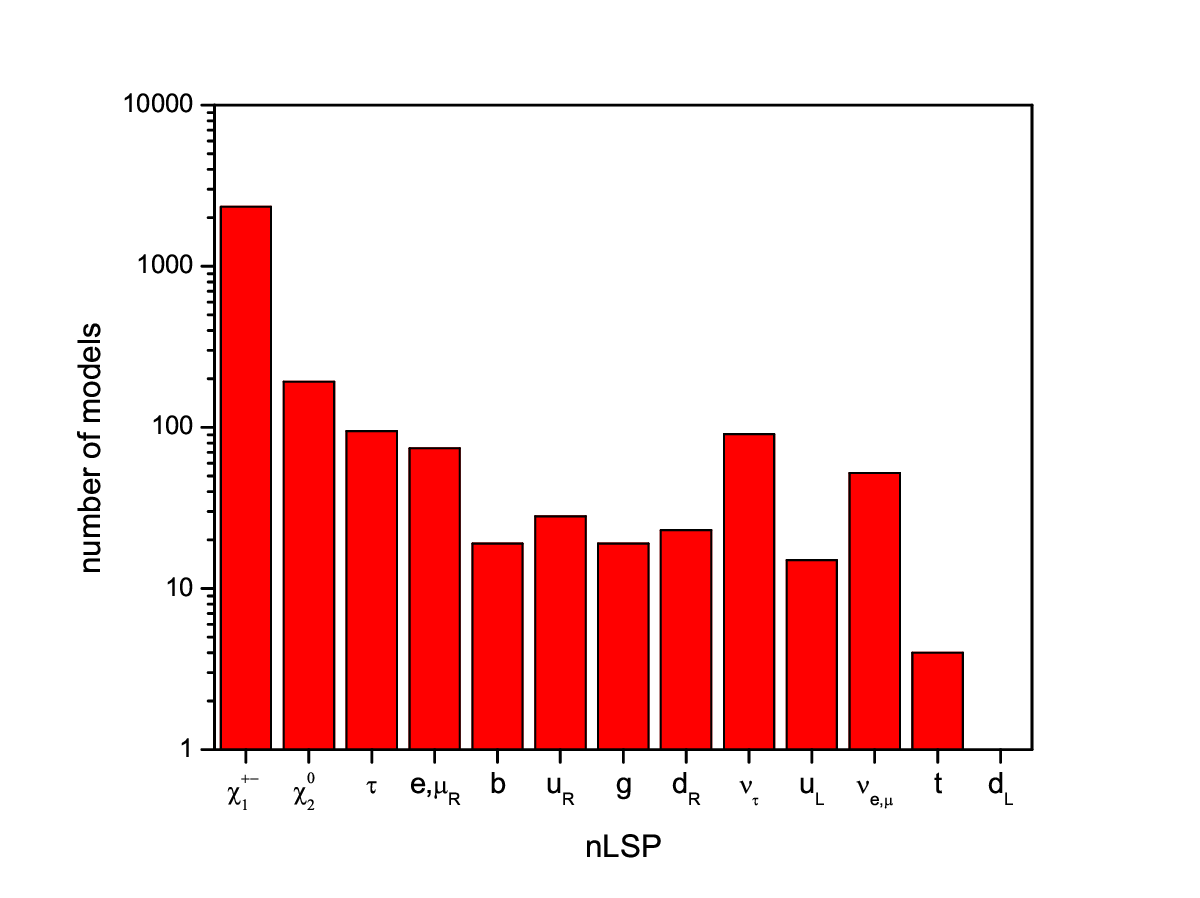}}
\caption{Identity of the nLSP in the case of flat(top) and log(bottom) priors.}
\label{Fig3}
\end{figure}

\begin{figure}[htbp]
\centerline{
\includegraphics[width=14.0cm,angle=0]{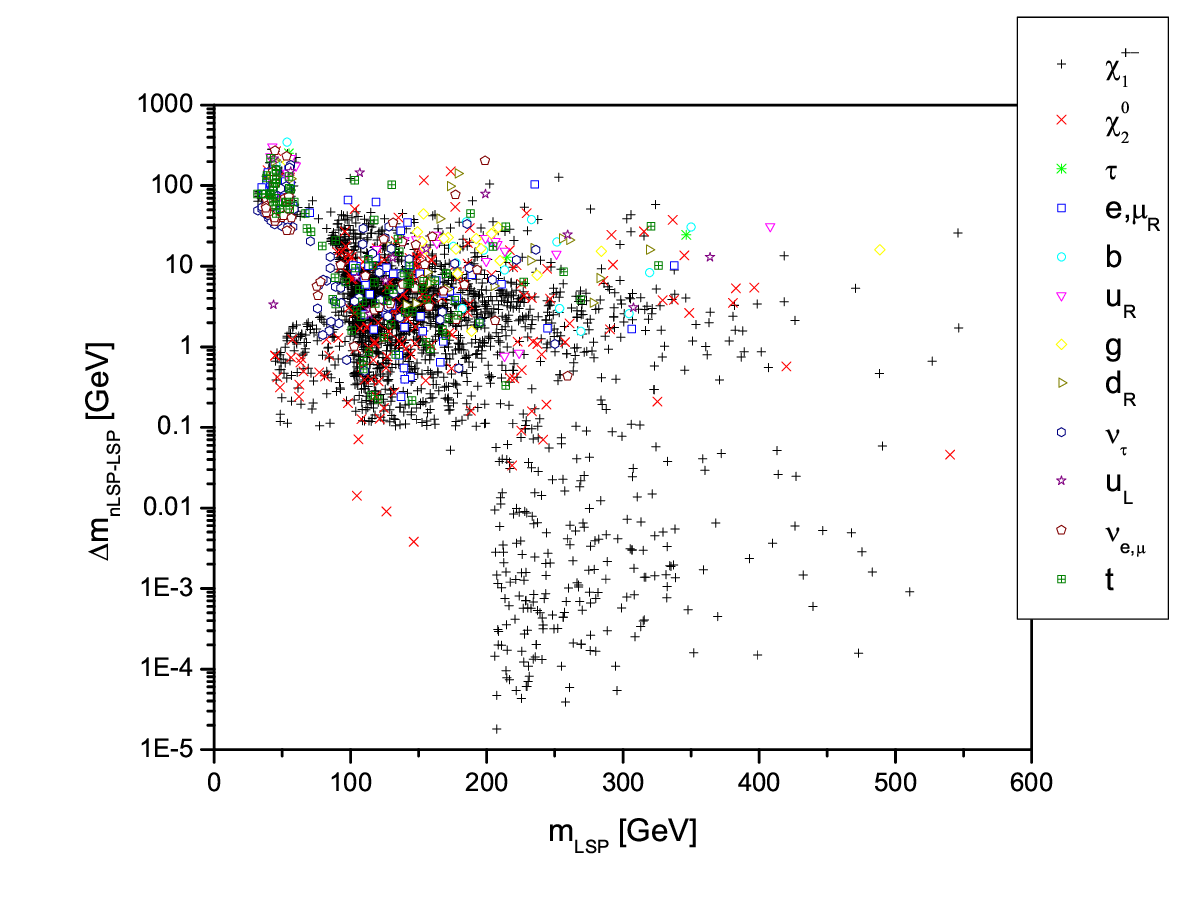}}
\caption{LSP-nLSP mass splitting as a function of the LSP mass showing the nLSP identity in the case of log priors.}
\label{Fig4}
\end{figure}

The next most important result for LC is the nature of the nLSP. Fig.~\ref{Fig3} shows that while $\tilde \chi_1^\pm$ and 
$\tilde \chi_2^0$ are seen to be the most likely nLSP candidates we find that almost any other sparticle can play this role 
with a probability on the order of a few percent for either choice of prior. Note that light right-handed squarks may be the 
nLSP in some cases; if the corresponding mass splitting with the LSP is small, \eg, less than a few GeV, these squarks will 
be very difficult to observe at the LHC and even at LC due to large $\gamma \gamma$ backgrounds.  We can see this explicitly 
in Fig.~\ref{Fig4} which shows the nLSP-LSP mass splitting for the different nLSP possibilities for the case of log priors.
With charginos playing the nLSP role we see the nLSP-LSP mass splitting can be particularly small and that most of the light 
chargino models in this case are removed by the Tevatron stable particle search as can be seen in the lower left-hand side of 
the figure.  

From this study it is clear that the MSSM can be much more complex than one would imagine by examining specific SUSY-breaking 
scenarios. If SUSY is discovered at the LHC it is quite clear that a LC will be needed in order for us to get a true picture 
of the underlying physics.

\begin{footnotesize}


\end{footnotesize}
\end{document}